\documentclass[a4paper,fleqn,usenatbib]{mnras}
\usepackage{newtxtext,newtxmath}
\usepackage[T1]{fontenc}
\usepackage{ae,aecompl}

\usepackage{graphicx}	% Including figure files
\usepackage{amsmath}	% Advanced maths commands
\usepackage{amssymb}	% Extra maths symbols
\usepackage{subfig}
\title[Flux distribution of individual blazars]{The flux distribution of individual blazars as a key to understand the dynamics of particle acceleration}

\author[A. Sinha et al.]{
Atreyee Sinha$^{1,2}$\thanks{E-mail: asinha@apc.in2p3.fr},
Rukaiya Khatoon$^{3}$\thanks{E-mail: rukaiyakhatoon12@gmail.com},
Ranjeev Misra$^{1}$, Sunder Sahayanathan$^{4}$,\newauthor Soma Mandal$^{5}$, Rupjyoti Gogoi$^{3}$ and Nilay Bhatt$^{4}$ 
\\
$^{1}$Inter-University Center for Astronomy and Astrophysics, Post Bag 4, Ganeshkhind, Pune-411007, India. \\
$^{2}$Now at: APC, AstroParticule et Cosmologie, Université Paris Diderot, CNRS/IN2P3, CEA/Irfu, Observatoire de Paris,\\
Sorbonne Paris Cité, 10, rue Alice Domon et Léonie Duquet, 75205 Paris Cedex 13, France.\\
$^{3}$Tezpur University,Napaam-784028, Assam, India.\\
$^{4}$Astrophysical Sciences Division, Bhabha Atomic Research Centre, Mumbai - 400085, India.\\
$^{5}$Government Girls' General Degree College, Kolkata-700023, West Bengal, India.
}

\date{Accepted XXX. Received YYY; in original form ZZZ}

\pubyear{2018}

\begin{document}
\label{firstpage}
\pagerange{\pageref{firstpage}--\pageref{lastpage}}
\maketitle

\begin{abstract}
The observed log-normal flux distributions in the high energy emission from blazars have been interpreted as being due to variability stemming from non-linear multiplicative processes generated dynamically from the accretion disc. On the other hand, rapid minute scale variations in the flux point to a  compact emitting region inside the jet, probably disconnected from the disc. In this work, we show that linear Gaussian variations of the intrinsic particle acceleration or escape time-scales can produce distinct non-Gaussian flux distributions, including log-normal ones. Moreover, the spectral index distributions can provide confirming evidence for the origin of the variability. Thus, modelling of the flux and index distributions can lead to quantitative identification of the micro-physical origin of the variability in these sources. As an
example, we model the X-ray flux and index distribution of Mkn~421 obtained from $\sim 9$ years of 
MAXI observations and show that the variability in the X-ray emission is driven by Gaussian fluctuations of the particle acceleration process rather than that of the escape rate.

\end{abstract}


%%%%%%%%%%%%%%%%% BODY OF PAPER %%%%%%%%%%%%%%%%%%

\section{Introduction}

Blazars are a special class of radio-loud active galactic nuclei (AGNs) and their observed broadband spectra 
are dominated by non-thermal emission arising from radiative cooling of relativistic electron distributions 
in powerful Doppler-boosted jets \citep{1995PASP..107..803U}. Additionally, blazar luminosity is observed to
vary over time scales of years down to minutes and at all wavelengths 
across the electromagnetic spectrum. Despite many decades of observations, the cause of the underlying variability 
is poorly understood. The dominance of the non-thermal emission further hinders our understanding of the accretion disk-jet
connection in these sources.

Irrespective of the origin, emission from blazars has been found to be 
stochastic in nature, similar to that seen in other AGNs and Galactic X-ray 
binaries \citep{McHardy,ritaban,Nakagawa,Sobolewska}. 
Since the past decade, much work has been done to understand the flux distribution of the lightcurves. For a linear stochastic process, a Gaussian distribution of the flux is to be expected, with the width of the distribution determining the flux variation. 
However, for the case of the eponymous blazar BL Lac, a log-normal flux distribution was clearly evident in the long-term X-ray 
light curves, with the average amplitude of variability being proportional to the flux level \citep{giebels_lognorm}.
Henceforth, this behaviour has been witnessed even in other blazars, and at different timescales and wavelengths \citep{2010A&A...520A..83H,my421, my1011, 2015ICRC...34..829C, pankaj_ln,2018arXiv180504675S}. Such properties were initially observed in the X-ray emission of the galactic black hole binary Cygnus X-1 \citep{lognorm_xrb}, and are usually interpreted as arising from multiplicative processes which originate in the 
accretion disc \citep{lyub,2005MNRAS.359..345U,2010LNP...794..203M}. 
However, minute time scale variability as seen in many 
blazars \citep{1996Natur.383..319G, 2007ApJ...664L..71A, vhe501, vaidehi421} is difficult to originate 
from the disc \citep{narayan}, and strongly favors the variability to originate within the jet. 

On the other hand, additive processes can also result in such distributions under specific scenarios.
\cite{minijet} studied the statistical properties of the mini jets-in-a-jet model of \cite{giannios} and 
 showed that the total flux from randomly oriented mini jets will converge to an 
$\alpha$-stable distribution. Further, inclusion of experimental uncertainties can imitate 
such a distribution as a log-normal one.
In this work, we provide an alternate interpretation of the non-Gaussian signatures seen in blazar variability through 
linear fluctuations of the underlying particle acceleration and/or the diffusive escape rate of the emitting electrons. 
Such small Gaussian perturbations propagate to produce non-linear flux distributions and linear flux-rms relations at 
high frequencies. This can explain the log-normal behaviour in both the long term stationary time series and 
during blazar flares, while reproducing the observed flux-rms relations. Finally, this study is used to interpret a 
plausible cause of variability in light-curves obtained from the MAXI observations for the brightest TeV blazar, Mkn 421.

\section{Perturbation on the intrinsic time scales}
We consider a scenario where the non-thermal electrons responsible for the blazar emission are accelerated at a shock front (AR; the acceleration region). Subsequently, they diffuse downstream (CR; the cooling region), at a rate ${\tau_{e}}^{-1}$, where they radiate through synchrotron and inverse Compton (IC) mechanisms \citep{kirk,2008MNRAS.388L..49S}
The kinetic equation describing evolution of the electrons in the AR can be written as \citep{kardashev},
\begin{equation}
	\label{eq:arkinetic}
	\frac{\partial n(\gamma,t)}{\partial t}+{\frac{\partial}{\partial \gamma}} \left[\left(\frac{\gamma}{\tau_{a}}-A\gamma^2\right)n(\gamma,t)\right]+\frac{n(\gamma,t)}{\tau_{e}}= Q(\gamma)   
\end{equation}
where, $\gamma/{\tau_{a}}$ is the electron acceleration rate and $A\gamma^2$ is the radiative loss 
rate .\footnote{Here, the radiative loss includes both synchrotron and inverse 
Compton processes happening at Thomson regime. For high electron and target photon   
energies the Compton scattering process will happen at Klein-Nishina regime and the loss rate
will be different. However, here we confine or study within the low energy
domain where Thomson approximation is valid.}
Together, they govern the maximum attainable Lorentz factor of the accelerated electrons, 
$\gamma_{max}=\frac{1}{{A}{\tau_{a}}}$. 
The steady state solution of  equation~(\ref{eq:arkinetic}) for a mono energetic electron injection, 
$Q(\gamma) = Q_0 {\delta({\gamma}-{\gamma}_{0})}$, will be
\begin{equation}
	\label{eq:steadyar}
	{n_0(\gamma)}= {Q_{0}}{\tau_{a}}{\gamma^{-1-\frac{\tau_{a}}{\tau_{e}}}}{\left(1-\frac{\gamma}{\gamma_{max}}\right)}^{\frac{\tau_{a}}{\tau_{e}}-1}{\left(\frac{1}{\gamma_0}-\frac{1}{\gamma_{max}}\right)^{-\frac{\tau_{a}}{\tau_{e}}}}
\end{equation}

After injection into the CR, the evolution of these particles is governed by
\begin{equation}
	\frac{\partial n_{s}(\gamma,t)}{\partial t}=\frac{\partial}{\partial \gamma} [B{\gamma^2}n_{s}(\gamma,t)]+\frac{n(\gamma,t)}{\tau_{e}}-\frac{n_{s}(\gamma,t)}{t_{e}}
	\label{eq:crkinetic}
\end{equation}
\noindent
where first term on the right hand side of equation (\ref{eq:crkinetic}) describes the radiative loss rate 
in the CR, and the last term is the escape of electrons from CR at a rate ${t_{e}}^{-1}$.
The steady state solution of the above equation will be a broken power law, 
with indices ${\tau_{a}}/{\tau_{e}}+1$ and ${\tau_{a}}/{\tau_{e}}+2$, and a break at 
energy ${1}/{B t_{e}}$. Since the indices of the particle spectrum do not depend on the intrinsic 
timescales of the CR, this will not introduce any additional non-linearity in the temporal behaviour. 
Moreover, as the radiative loss rate is $ \propto \gamma^2$, the resultant photon spectrum will 
again be a broken power law with indices ${\tau_{a}}/{2\tau_{e}}$ and 
${(\tau_{a} + \tau_{e})}/{2\tau_{e}}$ respectively. 
The narrow width of the single particle emissivity 
due to synchrotron and inverse Compton emission mechanisms, with respect to the power law electron 
distribution, further ensures that the photon spectrum will retain the temporal behaviour of the underlying particle distribution. In addition, the shape of the flux distribution due to 
synchrotron and inverse Compton scattering of an external photon field will be 
similar to that of electron number density since the corresponding emissivities are proportional to the 
number density. On the other hand, for synchrotron self Compton
process, the emissivity will depend on the square of the electron distribution  \citep{2018RAA....18...35S}
and hence the variance of the distribution will be twice as that of the electron distribution.

\subsection{Gaussian perturbation on ${\tau_{a}}$} \label{sec:tau_a_pert}
A small perturbation in the acceleration time scale can introduce variation in the accelerated particle
number density. If we quantify this variation in $\tau_a$ as
\begin{equation}
{\tau_a} =\tau_{a0}+\Delta\tau_a
	\label{eq:varyta}
\end{equation}
where, $\tau_{a0}$ corresponds to the mean acceleration timescale, the change in the number density
can be expressed as
\begin{equation}
	\bar{n}(\gamma)={\bar{n}_{0}(\gamma)}+{\Delta{\bar{n}}(\gamma)}
	\label{eq:varyn}
\end{equation}
where, $\bar{n}_{0}$ is the steady state solution (equation (\ref{eq:steadyar})) corresponding to $\tau_a=\tau_{a0}$.
Substituting equations~(\ref{eq:varyta}) and (\ref{eq:varyn}) in the steady state form of equation~(\ref{eq:arkinetic}), 
the fractional variability in $\bar{n}(\gamma)$ can then be obtained as, 
\begin{align}\label{eq:nvar}
	{\frac{\Delta{\bar{n}(\gamma)}}{\bar{n}(\gamma)}} &= f(\gamma) \frac{\Delta{\tau_a}}{\tau_{a}} + g(\gamma) \frac{\Delta \tau_a}{\tau_{e}}
\end{align}
where,
\begin{align}
	f(\gamma) &=\left(\frac{1}{1-\gamma/\gamma_{max}}\right)\\
\label{eq:ggamma}	
	g(\gamma) &= \log\frac{{\gamma_{0}}(1-\gamma/\gamma_{max})}{{\gamma}(1-\gamma_{0}/\gamma_{max})} - \frac{\gamma/\gamma_{max}}{1-\gamma/\gamma_{max}} + \frac{\gamma_{0}/\gamma_{max}}{1-\gamma_{0}/\gamma_{max}}
\end{align}
From equation (\ref{eq:nvar}), it is evident that the variability in $\bar{n}(\gamma)$ is a 
linear combination of Gaussian and log-normal 
terms. The relative amplitudes of these terms are decided by the functions $f(\gamma)$ and $g(\gamma)$. For the case 
$\gamma_{max}\rightarrow \infty$, the log-normal term dominates when $\gamma\gg\gamma_0\,\rm{exp}(\tau_{e}/\tau_a)$.
%{\bf Further in this case, the standard deviation of a normally 
%	distributed $\tau_a$ will be approximately $\tau_e/\rm{log}(\gamma_0/\gamma)$ times that of log$\;{\bar{n}(\gamma)}$. Since
%$\Delta \tau_a/\tau_e$ is related to the variability in spectral index and log$\;{\bar{n}(\gamma)}$ to
%logarithm of photon flux, the standard deviation of latter will be log$(\gamma_0/\gamma)$ times of the former in case 
%of synchrotron and external Compton processes.}
Also in this case, the standard deviation of a normally 
distributed $\tau_a$ will be approximately $\tau_e/\rm{log}(\gamma_0/\gamma)$ times that of log$\;{\bar{n}(\gamma)}$. 
Since the variability in photon index will be equal to $\Delta \tau_a / 2 \tau_e$ , the standard deviation of the 
logarithm of the photon flux distribution will be 2|log$(\gamma_0/\gamma)$| times the index distribution 
(in case of synchrotron and external Compton processes).

To quantify the deviation of $n(\gamma,t)$ from a Gaussian, we simulate its temporal behaviour by solving 
 equation (\ref{eq:arkinetic}) numerically using finite difference scheme. Gaussian perturbations 
of varying widths ($\sigma_{\tau_a}$) are then introduced in $\tau_a$ and the time series spanning over 
5000 points of $n(\gamma)$ is computed for each case at different values of $\gamma$. The values of
$\gamma_0$ and $\gamma_{max}$ are kept fixed at $10$ and $10^5$ respectively. 
The generated time series are then investigated for various statistical properties.

%\begin{figure*}
\begin{figure}
\centering
\includegraphics[scale=0.35,angle=-90]{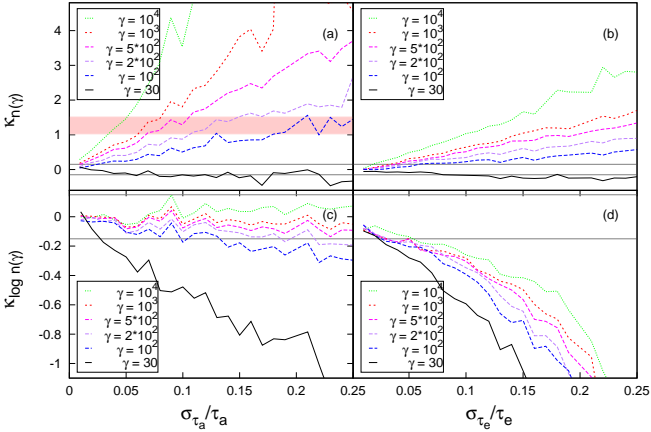}
\caption{Skewness ($\kappa$) of the simulated particle distribution as a function of $\sigma_{\tau_a}/\tau_a$  and $\sigma_{\tau_e}/\tau_e$ 
is shown in (a) and (b), whereas the skewness of the logarithm of the distribution as a function of $\sigma_{\tau_a}/\tau_a$ and 
$\sigma_{\tau_e}/\tau_e$ is shown in (c) and (d). 
The solid line corresponds to electron with Lorentz factor $\gamma = 30$ (black), dashed
line to $\gamma=10^2$ (blue), dashed
line to $\gamma=2*10^2$ (purple), dashed
line to $\gamma=5*10^2$ (magenta), short dashed line to $\gamma = 10^{3}$ (red) and dotted line $\gamma = 10^{4}$ (green). 
The grey lines show the 3-$\sigma$ ($3\sqrt{15/N}$) error range \citep{1992nrfa.book.....P}. 
The pink band shows the 1-$\sigma$ error range for the
observed value for Mkn421 (\S \ref{sect:disc}).} 
\label{fig:skewg}
\end{figure}
%\end{figure*}

\begin{figure}
\centering
\includegraphics[scale=0.35,angle=-90]{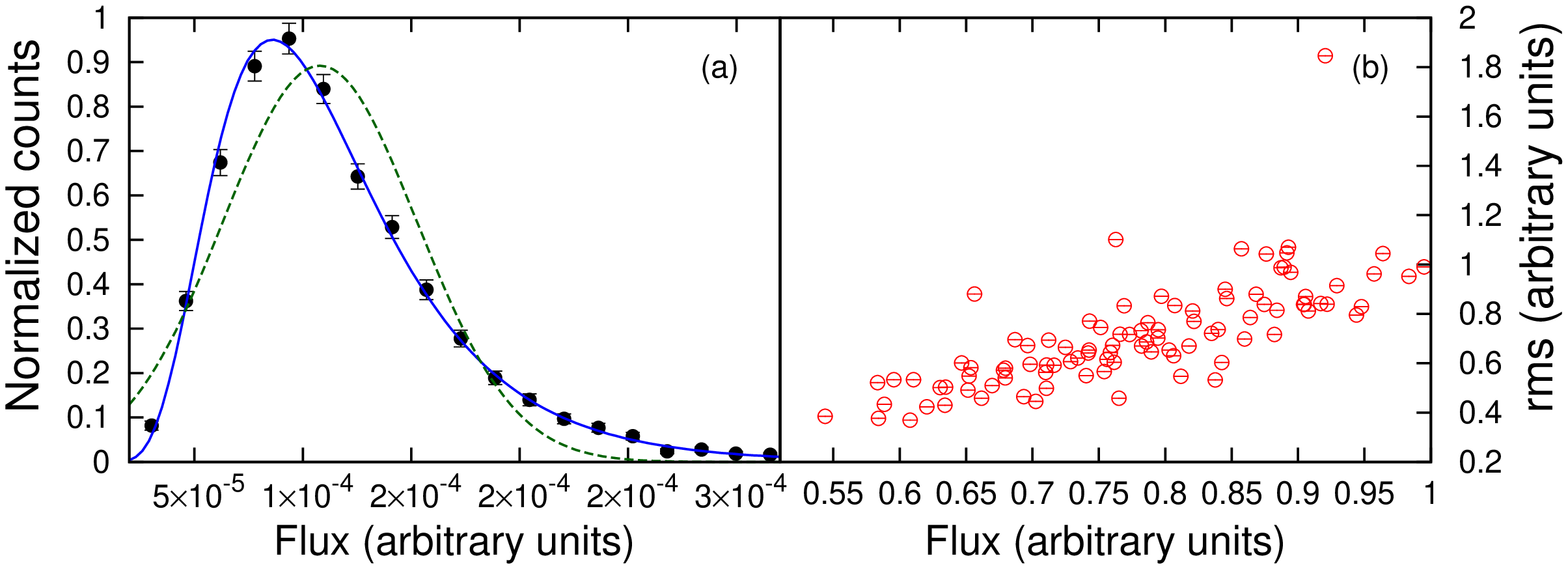}
\caption{(a) Histogram of the simulated particle number density for $\sigma_{\tau_a}/\tau_a = 0.1$ and $\gamma = 10^{3}$. 
The dashed green line represents the best fit Gaussian and the solid blue line represents the best fit log-normal PDF.
	(b) The flux-rms scatter plot obtained by dividing the simulated time series into 50 equal time bins. A strong positive correlation is clearly evident ($\rho = 0.83$, $P = 4\times10^{-26}$). }
\label{fig:flux_rms_ta}
\end{figure}

In Figure {\ref{fig:skewg}{(a)}}, we plot the skewness of the accelerated electron distribution ($\kappa_{n}$) as a function 
of $\sigma_{\tau_a}/\tau_a$ for different values of $\gamma$.  Since the fractional variation in 
the blazar spectral index during different flux states is 15\% approximately (see \S \ref{sect:disc}),
we extend $\sigma_{\tau_a}/\tau_a$ variation up to 0.25.  
At low electron energies ($\gamma \approx 30$) the skewness of the
distribution is negligible indicating a symmetric distribution. However, the distributions drift towards highly tailed ones 
for increasing electron energies, thus implying a deviation from Gaussianity. To investigate whether the skewed high 
energy electron distribution reflects a log-normal behaviour, we plot in Figure {\ref{fig:skewg}{(c)}}, the skewness of the 
logarithm of the number density ($\kappa_{{\rm log}\,n}$) as a function of $\sigma_{\tau_a}/\tau_a$. Here, the skewness
is negligible for increasing electron energies suggesting a possible drift towards a log-normal distribution. 
To confirm this, we further fit the normalised distribution of the number densities 
with Gaussian and log-normal probability density functions (PDFs). We find that a log-normal PDF significantly fits
the distribution better at high 
electron energies. In Figure {\ref{fig:flux_rms_ta}{(a)}}, we show the normal and 
log-normal fit to the electron distribution corresponding to $\gamma=10^3$ and $\sigma_{\tau_a}/\tau_a=0.1$. 
Clearly, the fit statistics is better for a log-normal with a reduced chi-square, $\chi_{\rm red}^2 \approx 1.1$ for 17 degrees of freedom (dof), 
than a Gaussian ($\chi_{\rm red}^2 \approx 20.9$ for 17 dof) PDF. 
The log-normal behaviour of the number density $n$ at large $\gamma$ lets us express the skewness of the 
distribution as  
\begin{align}\label{eq:skewln}
	\kappa_n=\left(2+e^{\sigma_{\rm{log}\,n}^2}\right)\,\sqrt{e^{\sigma_{\rm{log}\,n}^2}-1}
\end{align}
where, $\sigma_{\rm{log}\,n}$ is the standard deviation of log $n$ which can be approximated as
\begin{align}\label{eq:sigln}
	\sigma_{\rm{log}\,n} \approx \frac{g(\gamma)}{\tau_{e}}\sigma_{\tau_a}
\end{align}
The energy dependence of $\sigma_{\rm{log}\,n}$ will cause the skewness ($\kappa_n$) to increase with energy which in turn can be an 
indicator for the energy of the emitting electrons. It is evident from equations (\ref{eq:skewln}) and (\ref{eq:sigln}) that for 
$\sigma_{\tau_a}\rightarrow 0$, the distribution of $n$ will closely reflect a Gaussian behaviour.
  
\begin{figure}
        \centering
 {\includegraphics[scale=0.35,angle=-90]{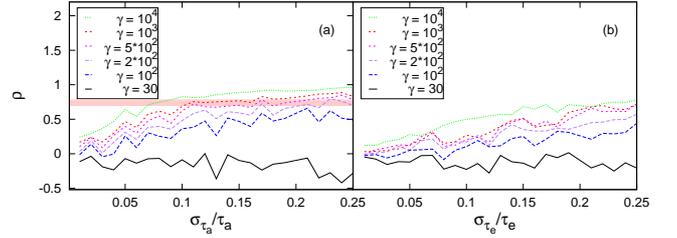}
	\caption{Spearman's rank correlation coefficient $\rho$ of the flux-rms scatter plot as a function of (a) $\sigma_{\tau_a}/\tau_a$ and (b)
$\sigma_{\tau_e}/\tau_e$. 
%Different lines correspond to electron with Lorentz factors $\gamma = 30$, $10^2$, $10^{3}$ and $10^{4}$. 
The legends 
are same as in Figure \ref{fig:skewg}. The pink band shows the 1-$\sigma$ ($0.6325/\sqrt{N-1}$) error range for the
observed value for Mkn421 (\S \ref{sect:disc}).} 
	 \label{fig:corr_siga}}

\end{figure}

A necessary feature of a log-normal behaviour is a linear dependence of the average flux on its excess (rms) variation \citep{2003MNRAS.345.1271V}.
Consistently, the electron number density at high energies should reflect this behaviour and to examine this,
we compute the average number density and its variation, for a given $\gamma$ and $\sigma_{\tau_a}/\tau_a$, 
by dividing the corresponding time series into 50 equal time bins. In Figure \ref{fig:flux_rms_ta}{(b)}, we show the distribution of
the average number density and its variation for $\gamma=10^3$ and $\sigma_{\tau_a}/\tau_a =0.1$.
A Spearman's rank correlation study shows these quantities are significantly correlated with correlation
coefficient $\rho = 0.83$ with null hypothesis probability $P = 4\times10^{-26}$. In Figure \ref{fig:corr_siga}{(a)}, we plot the 
correlation coefficient with respect to $\sigma_{\tau_a}/\tau_a$ for different
values of $\gamma$. It can be noted that the correlation improves with the increasing value of $\gamma$, thereby supporting a log-normal behaviour. 

\begin{figure}
\qquad
\centering
{\includegraphics[scale=0.35,angle=-90]{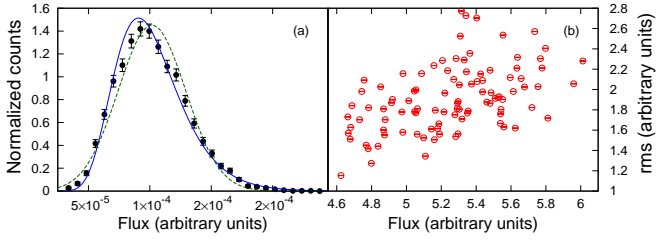}
\caption{(a) Histogram of the simulated particle number density for $\sigma_{\tau_e}/\tau_e = 0.1$ and $\gamma = 10^{3}$. 
	The dashed green line represents the best fit Gaussian and the solid blue line, the best fit log-normal PDF.
	(b) The flux-rms scatter plot obtained by dividing the simulated time series into 50 equal time bins. A weak positive correlation is seen ($\rho = 0.26$, $P = 0.0071$). }
\label{fig:flux_rms_te}}
\end{figure}

\subsection{Gaussian perturbation on ${\tau_{e}}$}

In addition to the acceleration rate, the observed photon spectral index will also depend on the confinement time of the 
electron distribution within AR. In other words, a variation in the escape timescale in AR can introduce non-linearity in 
the electron distribution. To study this effect, we quantify the 
variation in escape time scale (${\tau_{e}}$) in AR as 
\begin{equation}
	{{\tau_{e}}} ={\tau_{e0}+{\Delta{\tau_{e}}}}
	 \label{eq:varytea}
\end{equation} 
\noindent
and the corresponding change in the electron number density as 
\begin{equation}
\tilde{n}(\gamma)={\tilde{n}_{0}(\gamma)}+{\Delta{\tilde{n}}(\gamma)}
     \label{eq:varynn}
\end{equation}
\noindent
where, $\tilde{n}_{0}$ is the steady state solution (equation (\ref{eq:steadyar})) corresponding to $\tau_{e}=\tau_{e0}$.
Following the procedure similar to the case of $\tau_a$ (\S \ref{sec:tau_a_pert}), substituting equations~(\ref{eq:varytea}) and (\ref{eq:varynn}) 
in the steady state form of equation~(\ref{eq:arkinetic}), the fractional variability 
in $\tilde{n}(\gamma)$ can then be obtained as,
\begin{align}
\label{eq:ntilde}
	\frac{\Delta \tilde{n}}{\tilde{n}} = \tau_a \frac{\Delta \tau_{e}}{\tau_{e}^2} f(\gamma)
\end{align}
where,
\begin{align}
    f(\gamma) =& \log\frac{\gamma(1-\gamma_0/\gamma_{max})}{\gamma_0(1-\gamma/\gamma_{max})} 
\end{align}
It is evident from equation (\ref{eq:ntilde}), that while the resultant distribution will be neither
normal nor log-normal, it will be a skewed one.
Additionally, since the particle index $p \sim \frac{\tau_{a}}{\tau_{e}}$, the distribution of the 
spectral indices will also be skewed. 

To further quantify the effect on the electron number density due to a Gaussian 
fluctuation in $\tau_{e}$, we simulate  the 
temporal behaviour of $n(\gamma,t)$ by solving equation (\ref{eq:arkinetic}) numerically (\S \ref{sec:tau_a_pert}).
In Figure \ref{fig:skewg}{(b)}, we show the skewness of the particle distribution ($\kappa_{n})$ as a function 
of $\sigma_{\tau_e}/\tau_{e}$ for different values of $\gamma$. The distributions are highly 
tailed for increasing values of $\gamma$ supporting a non-Gaussian behaviour.  A similar behaviour 
is also observed in case of the skewness of the logarithm of the number density  ($\kappa_{\rm{log}\,n}$) which is 
shown in Figure \ref{fig:skewg}{(d)}.
These studies suggest that the resultant electron number density distribution is significantly skewed; however, it is 
neither normal nor log-normal. 

We also perform the Anderson Darling test on the distribution of the electron number
density for various $\gamma$. Consistent with our earlier study, both Gaussian and log-normal fits are 
strongly rejected. In Figure \ref{fig:flux_rms_te}{(a)}, we show the normalised histogram of the electron 
number density for $\gamma = 10^3$ and $\sigma_{\tau_e}/\tau_{e}$ = 0.1 fitted with Gaussian and log-normal PDFs. 
Our fit result suggests both of these PDFs cannot represent the given 
distribution with $\chi_{\rm red}^2 \approx 6.7$ (dof=28) 
for the Gaussian PDF and $\chi_{\rm red}^2 \approx 3.7$ (dof=28) for log-normal one. 
To study the flux-rms relation, we divide the 
temporal evolution of the number density into 50 equal time bins (\S \ref{sec:tau_a_pert}), the average number density 
and its rms variation corresponding to each bin is determined. In Figure \ref{fig:flux_rms_te}{(b)}, we show their 
distribution for the case of $\gamma= 10^3$ and $\sigma_{\tau_e}/\tau_{e}$ = 0.1. A Spearman's rank correlation study 
suggests mild positive correlation between these quantities with $\rho =0.26$ and $P=0.0071$. The variation of 
the flux-rms correlation coefficient with respect to $\sigma_{\tau_e}/\tau_{e}$ is shown in Figure \ref{fig:corr_siga}{(b)}
for different values of $\gamma$. The correlation improves with increasing value of $\gamma$; however, it is less significant 
than the case of the Gaussian perturbation on $\tau_a$.

\section{Discussion}\label{sect:disc}

%The linear behaviour in the relation of RMS and flux of individual blazars or the skewness shown by the distribution of the flux 
The flux-rms relation of individual blazars or the skewness shown by the distribution of the flux 
are interpreted by several authors as arising from multiplicative processes, favouring a variability stemming from the disk. 
Alternatively, \cite{minijet} demonstrated that such behaviour can also arise from a collection of
randomly oriented mini jets within the jet. They showed that the flux from a randomly oriented 
mini jet will follow a Pareto distribution which preserves the flux-rms relation. Further, the total 
flux due to several randomly oriented mini jets will be a sum of Pareto distributions that converge 
to an $\alpha$-stable distribution. The resultant flux distribution still holds the flux-rms 
relation; however, will neither be normal nor log-normal one. Nevertheless, inclusion of 
experimental uncertainties may imitate the distribution as a log-normal one.

\begin{figure}
\centering
    {\includegraphics[scale=0.35,angle=-90]{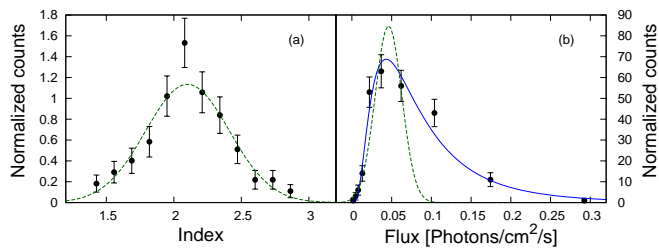}
	\caption{ Histograms of the 10 day binned X-ray (a) spectral index at 2-10 keV and (b) 2-20 keV flux of Mkn 421 spanning over 9 years. 
The dashed green line corresponds to the best fit Gaussian function, whereas the solid blue line corresponds to the best fit log-normal one.}
	  \label{fig:Mkn421}}
\end{figure}

In the present work, we show that small temporal fluctuations in the intrinsic time scales in the AR is capable of producing 
particle distributions with non-Gaussian signatures and significant flux-rms correlations. 
The novelty of this work is that it connects the long term temporal behaviour of the blazars with the
relatively shorter timescales of the acceleration process, and provide clues on electron energies responsible 
for the emission. To highlight this, we study the X-ray observations of the 
blazar Mkn\,421 by \emph{MAXI} satellite, spanning over 9 years ranging from 2009 to 2018 \citep{2009PASJ...61..999M}. While the integrated
counts obtained from a 10 days binned light curve showed a log-normal behaviour with $\chi_{\rm red}^2 \approx 1.43$ for 7 dof and ${\sigma = 0.33 \pm 0.02}$ over a Gaussian one with $\chi_{\rm red}^2 \approx 9.84$ for 7 dof (Figure \ref{fig:Mkn421}(b)), the spectral indices
estimated from the hardness ratio between 4--10 keV and 2--4 keV fluxes were normally 
distributed with $\chi_{\rm red}^2 \approx 0.81$ for 10 dof, mean $m_p = 2.1 \pm 0.022$ and standard deviation $\sigma_p=0.31 \pm 0.096$ (Figure \ref{fig:Mkn421}(a)).  
This suggests that the plausible physical process
responsible for the observed flux variation is associated with the fluctuations in the particle acceleration 
rate. 
%Since the X-ray emission from Mkn\,421 is attributed to synchrotron process, the similarity 
%between the standard deviations further confirm this inference.
The fractional variation in acceleration timescale can then be identified from $\sigma_p$ and $m_p$ of the index distribution as  $\sigma_{\tau_a}/\tau_a \approx 0.148\pm0.046\,$.
A comparison of Figure \ref{fig:skewg}(a) with this value and the observed skewness of $\kappa = 1.27\pm 0.24$ suggests the emission to originate from electrons with $\gamma$ range $\sim 10^2 - 10^3$. From Figure \ref{fig:corr_siga}(a), we see that this value of $\gamma$ is consistent with the observed correlation co-efficient $\rho = 0.74\pm 0.04$. However, this estimate of $\gamma$ is significantly lower than the electron energies obtained 
through the broadband spectral modelling of the source using synchrotron and inverse Compton emission 
mechanisms \citep{2016MNRAS.463.4481Z,my421,2011ApJ...736..131A,2009ApJ...691L..13D}. This discrepancy can be 
associated with the low value of the injection lorentz factor $\gamma_0$ which is fixed at $10$ for the present study. 
From equations (\ref{eq:ggamma}), (\ref{eq:skewln}) and (\ref{eq:sigln}), it is evident that the skewness is a function of 
$\gamma_0/\gamma$ rather than $\gamma$ alone. For a given $\kappa$, higher values of $\gamma_0$ can result in large
$\gamma$ values that may be consistent with the ones obtained through spectral modelling.

%However, this estimation is significantly lower than the electron energies obtained through the broadband spectral modeling of the source using synchrotron and inverse Compton emission mechanisms \citep{2016MNRAS.463.4481Z,my421,2011ApJ...736..131A,2009ApJ...691L..13D}. This discrepancy can be associated with the uncertainty in the spectral index and/or the parameters obtained through spectral modeling may not be well constrained. Specifically, since the index here is estimated from the hardness ratios, the spread in the indices is larger than what would be obtained from spectral fitting, leading to a higher estimation of $\sigma_{\tau_a}/\tau_a$, and in turn, lower values of $\gamma$. Moreover, the $\gamma$ range obtained in this work may be underestimated in case the injection lorentz factor $\gamma_0$ is significantly higher than the assumed value of 10. As we can see from equation (\ref{eq:ggamma}), (\ref{eq:skewln}) and (\ref{eq:sigln}),  $\kappa \sim f({\gamma_0}/{\gamma})$, and any scaling in $\gamma_0$ will change the final estimation of gamma accordingly.}

%This discrepancy can be associated with the  
%uncertainty in the parameters estimated here and/or through spectral modeling.}

While the log-normal distribution obtained from the light curve of blazars are generally integrated over a certain 
energy band, here we have quantified the distributions at some fixed electron energy. However, to be consistent with the 
observations, we verified our results for integrated number densities between different electron energies. 
We found that our results remain qualitatively similar to that obtained for the case of mono-energetic electron, being strongly dominated by the number counts at the lower energies.

\section{Conclusion}

Through the present work, we show that non-Gaussian flux distributions observed in blazars can be associated with 
the perturbations in the intrinsic time scales of the main particle acceleration region. A log-normal flux distribution
with the spectral indices showing a Gaussian behaviour can be attributed to the fluctuations in the acceleration rate,
whereas fluctuations in the electron escape rate can cause flux/index distributions which significantly differ from 
Gaussian and log-normal ones. Given well sampled multi-wavelength lightcurves, this study can be effectively utilized to identify the underlying physical processes, specifically in estimating the fractional fluctuations
in the intrinsic timescales and also the typical electron energies responsible for emission in the different frequency bands. In addition, by a comparison of the flux distributions at different energies
(e.g X-ray and gamma-ray), it is possible to identify whether the emission is associated with similar electron energies and thus, constrain spectral models.

\section*{Acknowledgements}
We are thankful to the anonymous referee for valuable comments and suggestions. S. Mandal is thankful to ISRO (ISRO/RES/2/404/15-16) and R. Khatoon is thankful to CSIR, New Delhi (03(1412)/17/EMR-II) for financial support. R. Gogoi would like to thank IUCAA, Pune for associateship.


\bibliographystyle{mnras}

%%%%%%%%%%%%%%%%% APPENDICES %%%%%%%%%%%%%%%%%%%%%

% Don't change these lines
\bsp	% typesetting comment
\label{lastpage}
\end{document}